\documentclass{article}
\usepackage[affil-it]{authblk}
\usepackage{graphicx}
\usepackage[space]{grffile}
\usepackage{latexsym}
\usepackage{textcomp}
\usepackage{longtable}
\usepackage{tabulary}
\usepackage{booktabs,array,multirow}
\usepackage{amsfonts,amsmath,amssymb}
\providecommand\citet{\cite}
\providecommand\citep{\cite}

\usepackage{url}
\usepackage{hyperref}
\hypersetup{colorlinks=false,pdfborder={0 0 0}}
\usepackage{etoolbox}
\makeatletter
\patchcmd\@combinedblfloats{\box\@outputbox}{\unvbox\@outputbox}{}{%
  \errmessage{\noexpand\@combinedblfloats could not be patched}%
}%
\makeatother
\newif\iflatexml\latexmlfalse
\AtBeginDocument{\DeclareGraphicsExtensions{.pdf,.PDF,.eps,.EPS,.png,.PNG,.tif,.TIF,.jpg,.JPG,.jpeg,.JPEG}}

\usepackage[utf8]{inputenc}
\usepackage[ngerman,english]{babel}

\begin{document}

\title{Generation of Shear Waves by Laser in Soft Media in the Ablative and Thermoelastic Regimes.}

 \author{Pol Grasland-Mongrain$^{1}$, Yuankang Lu$^{1,2}$, Frederic Lesage$^{2,3}$, Stefan Catheline$^{4}$, Guy Cloutier$^{1,3,5}$\\
\textit{(1) Laboratory of Biorheology and Medical Ultrasonics, Montreal Hospital Research Center, Montreal (QC), H1X0A9, Canada\\
(2) Departement of Electrical Engineering, École Polytechnique of Montreal, Montreal (QC), H3C3A7, Canada\\
(3) Institute of Biomedical Engineering, École Polytechnique and University of Montreal, Montreal (QC), H3T1J4, Canada\\
(4) Laboratory of Therapeutic Applications of Ultrasound, Inserm u1032, Inserm, Lyon, F-69003, France\\
(5) Departement of Radiology, Radio-oncology and Nuclear Medicine, University of Montreal, Montreal (QC), H3C3J7, Canada}}

\date{\today}

\maketitle

\selectlanguage{english}
\begin{abstract}
This article describes the generation of elastic shear waves in a soft medium using a laser beam. Our experiments show two different regimes depending on laser energy. Physical modeling of the underlying phenomena reveals a thermoelastic regime caused by a local dilatation resulting from temperature increase, and an ablative regime caused by a partial vaporization of the medium by the laser. Computed theoretical displacements are close to experimental measurements. A numerical study based on the physical modeling gives propagation patterns comparable to those generated experimentally. These results provide a physical basis for the feasibility of a shear wave elastography technique (a technique which measures a soft solid stiffness from shear wave propagation) by using a laser beam.%
\end{abstract}%

When a laser beam of sufficient energy is incident on a medium, the absorption of the electromagnetic radiation leads to an increase in the local temperature. Due to thermal effects, displacements occur in the medium, which can then propagate as elastic waves. Elastic waves within a bulk can be separated into two components: compression waves, corresponding to a curl-free propagation, and shear waves, corresponding to a divergence-free propagation \cite{aki1980quantitative}. Measures of the transmission characteristics of compression and shear waves are useful for inspecting solids, such as a metal, to reveal potential cracks or defects \cite{Shan_1993}. In biological tissues, induction of compression waves by laser has been studied with the development of photoacoustic imaging \cite{Xu_2006}. Elastic waves used in photoacoustic imaging are typically of a few megahertz; at this frequency, shear waves are quickly attenuated in soft tissues, typically over a few microns, and only compression waves can propagate over a few centimeters.

While the induction of surface acoustic waves by laser in soft tissues was recently demonstrated by Li et al. \cite{Li_2012}, \cite{Li_2014}, a similar phenomenon with shear waves in bulk medium has never been described. This is of great interest for the shear wave elastography technique. As its name indicates, shear wave elastography comprises the techniques used to map the elastic properties of soft media using shear wave propagation \cite{Plewes_1995}, \cite{muthupillai1995magnetic}, \cite{Catheline_1999}. These techniques typically use low frequency (50-500 Hz) shear waves so that their propagation can be observed over a few centimeters. The shear waves are currently generated using either an external shaker or a focused acoustic wave. However, alternative shear wave generation methods have drawn important attention recently. For example, it has been demonstrated that one can use natural motion of the medium \cite{Hirsch_2012}, \cite{Zorgani_2015}, the Lorentz force \cite{Basford_2005}, \cite{Grasland_Mongrain_2014}, \cite{Grasland_Mongrain_2016} or electrolysed-induced bubbles \cite{Montalescot_2016}. Compared to these generation sources, a laser presents the advantage of being fully remote, without need of coupling gel; and of being miniaturizable at low cost using an optical fiber.

\begin{figure}[h!]
\begin{center}
\includegraphics[width=1.00\columnwidth]{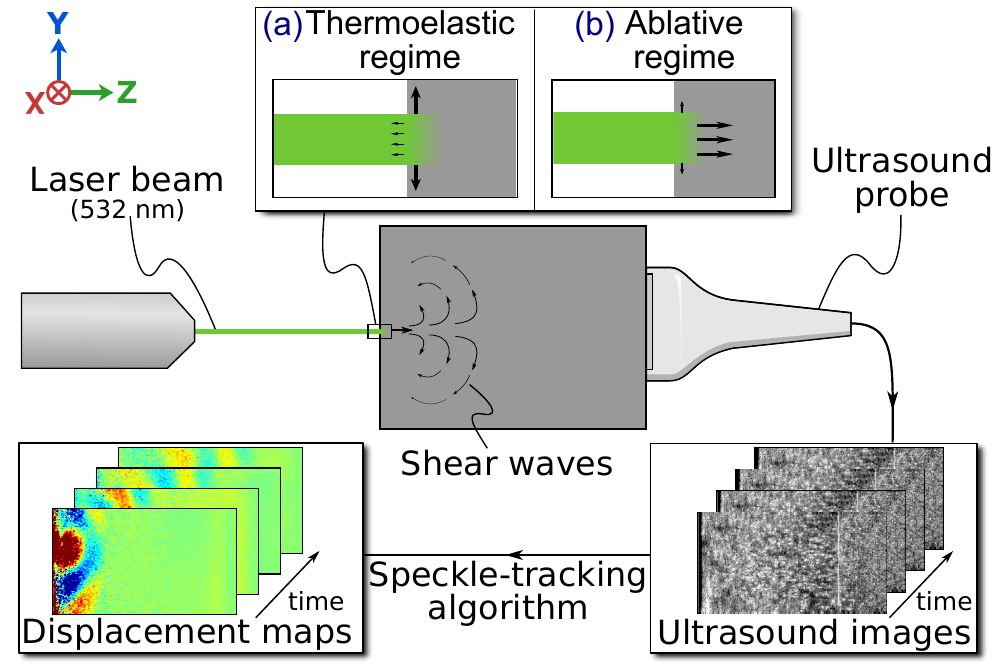}
\caption{{\label{Figure1} Experimental setup. A laser beam is emitted on a soft medium. This generates shear waves following (a) thermoelastic and/or (b) ablative regimes. The medium is observed with an ultrasound probe. A speckle-tracking algorithm calculates displacements from the ultrasound images.%
}}
\end{center}
\end{figure}

In our experiment, illustrated in Figure \ref{Figure1}, we used a Q-switch Nd:YAG laser (EverGreen 200, Quantel, Les Ulis, France) to produce a 10-ns pulse of 10 to 200 mJ energy at a central wavelength of 532 nm in a 5-mm diameter circular beam. The laser beam was absorbed in a 4x8x8 cm$^3$ black mat phantom composed of water, 5\% polyvinyl alcohol, and 1 \% black graphite powder. Two freeze/thaw cycles were applied to stiffen the material to a shear modulus of 25$\pm$5 kPa \cite{17375819}. To observe the resulting shear waves, the medium was scanned simultaneously with a 5-MHz ultrasonic probe, consisting of 128 elements, connected to a multi-channeled ultrasound scanner (Verasonics V-1, Redmond, WA, USA), and placed on the other side of the sample. The probe was acquiring 4000 ultrasound images per second during 30 ms. Due to the presence of graphite particles, the medium presented a speckle pattern on the ultrasound image.The computation of displacements along the Z axis (the ultrasound axis) was computed by tracking the speckle spots with the Lucas-Kanade method \cite{lucas1981iterative}. This method solves basic optical flow equations by least squares criterion in a window of 64*5 pixels centered on each pixel. Displacements over time were then filtered from 200 to 800 Hz using a 5th-order Butterworth filter and averaged over four experiments.\selectlanguage{english}

\begin{figure}[h!]
\begin{center}
\includegraphics[width=1.00\columnwidth]{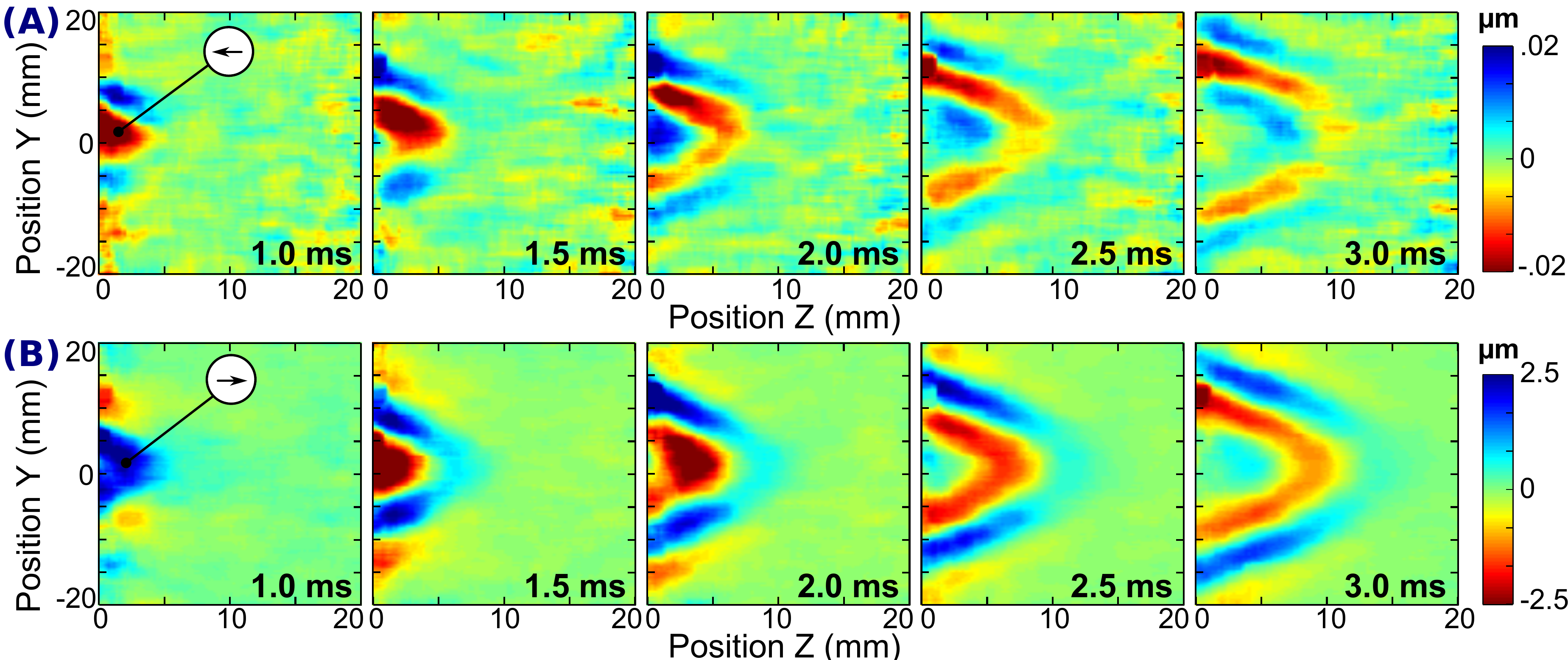}
\caption{{\label{figElastoPVA} Experimental displacement amplitude maps along the Z axis with a (A)-10 mJ or (B) 200-mJ laser beam, 1.0, 1.5, 2.0, 2.5 and 3.0 ms after laser emission in the tissue-mimicking phantom. In both cases, direction of the initial central displacements is indicated in the white circle.%
}}
\end{center}
\end{figure}

Figure \ref{figElastoPVA} illustrates the resulting displacement amplitude maps observed along the ultrasound axis at 1.0, 1.5, 2.0, 2.5, and 3.0 ms after laser emission for two laser beam energies (10 and 200 mJ). Displacements reached an amplitude of 0.02 $\mu$m for the 10-mJ laser beam and 2.5 $\mu$m for the 200-mJ laser beam. They propagated at a velocity of 5.5$\pm$0.5 m.s$^{-1}$, which is typical for a shear wave, but far slower than the usual velocity of a compression wave (about 1500 m.s$^{-1}$ in soft tissues). Shear modulus can then be calculated using the relationship $v_s = \sqrt{\mu/\rho}$. Supposing the medium density, $\rho$, at 1000 kg.m$^{-3}$ (water density), the propagation velocity corresponds to a shear modulus of 30$\pm$5 kPa, which is in the range of the expected value for this phantom. Shear wave frequency spectrum was centered at 500 $\pm$ 50 Hz.

Careful observation reveals several differences in the propagation patterns of the two laser beam energies. At low energy, first central displacements are directed towards the outside of the medium (left arrow), and three half cycles are observed. Conversely, at high energy, first displacements are directed towards the inside of the medium (right arrow), and only two half cycles can be observed.\selectlanguage{english}

Let's examine now the physical phenomena involved in these experiments. The optical intensity, $I_0$, of the laser beam is defined as $I_0=\frac{1}{S}\frac{d E}{dt}$, where $S$ is the beam surface and $E$ is the beam energy. When emitted on the surface of a medium and in the absence of reflection, the laser beam is absorbed with an exponential decay as a function of the medium depth $z$: $I(z)=I_0 \exp(- \gamma z)$, where $\gamma$ is the absorption coefficient of the medium. We have experimentally estimated $\gamma$ by measuring the fraction of light transmitted through different slices of the medium (with thicknesses of 0, 30, 50, and 100 $\mu$m) with a laser beam energy-measurement device (QE50LP-S-MB-D0 energy detector, Gentec, Qu\'ebec, QC, Canada). We found respective transmitted powers of 100\%, 42\%, 28\%, and 11\%, which gives $\gamma^{-1} \approx$ 40 $\mu m$ in our sample as indicated by an exponential fit.

Absorption of the laser beam by the medium subsequently gives rise to an absorbed optical energy, $\gamma I$. Assuming that all the optical energy is converted to heat, a local increase in temperature occurs. Temperature distribution, $T$, in the absence of convection and phase transition, can be computed using the following heat equation:
\begin{equation}
    k \nabla ^2 T = \rho C \frac{\partial T}{\partial t} - \gamma I
    \label{eq:eqChaleur}
\end{equation}
where $k$ is the thermal conductivity, $\rho$ is the density, $C$ is the heat capacity and $t$ is the time. Calculating the exact solution to this equation is beyond the scope of this article, but we can roughly approximate the first and second terms to be $k T / \gamma^2$ and $\rho C T / \tau$, respectively, during laser emission. Given that $k$ = 0.6 W.m$^{-1}$.K$^{-1}$ (water thermal conductivity), $\rho$ = 1000 kg.m$^{-3}$ (water density), $C$ = 4180 J.kg$^{-1}$.m$^{-3}$ (water heat capacity), $\gamma^{-1} \approx$ 40 $\mu$m and $\tau$ = 10 ns, the first term is negligible compared to the second one. Substituting low-energy experimental parameters ($E$ = 10 mJ, $S$ = 20 mm$^{2}$) leads to a maximum increase in temperature of 3 K, which produces a local dilatation of the medium. The induced displacements can then generate shear waves, which constitutes the \textit{thermoelastic regime}.

\begin{figure}[h!]
\begin{center}
\includegraphics[width=1.00\columnwidth]{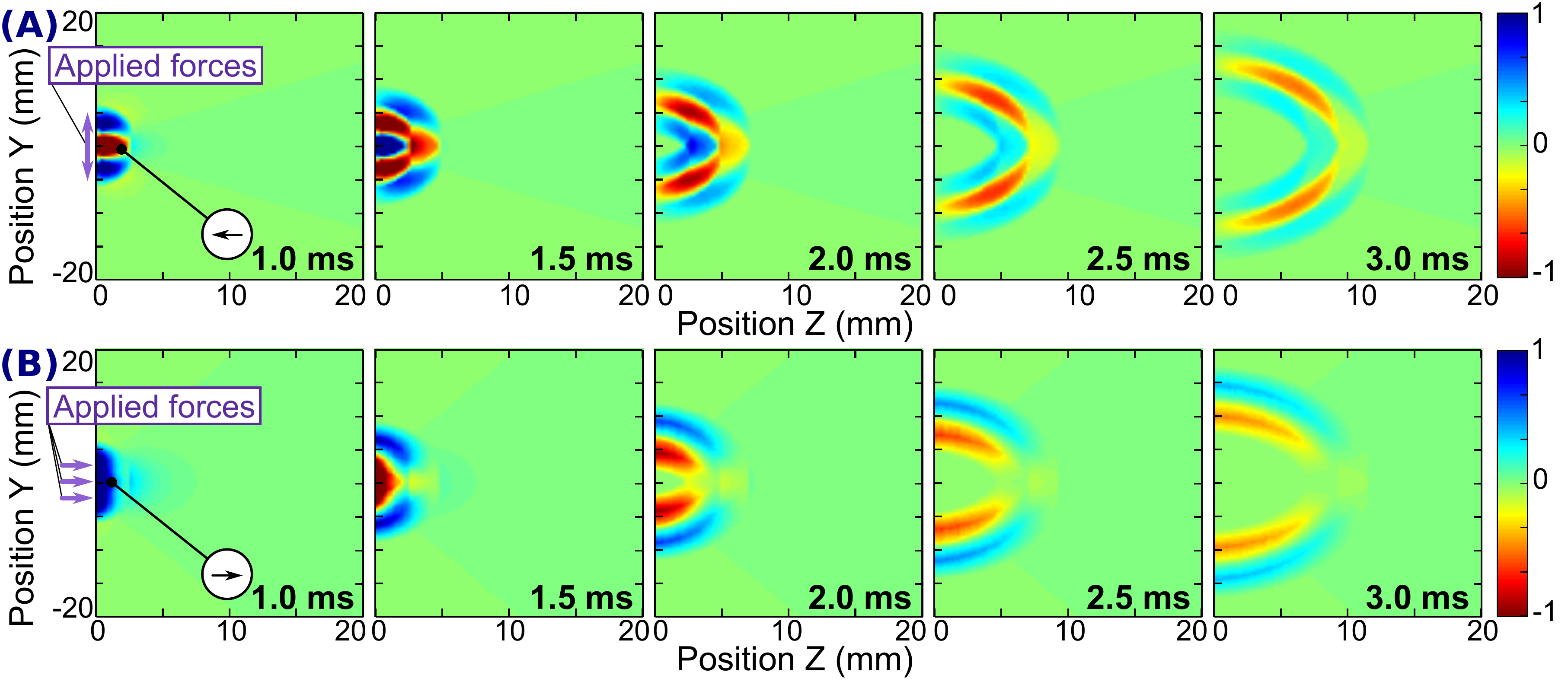}
\caption{{\label{figGreen} Normalized displacement maps along the Z axis 1.0, 1.5, 2.0, 2.5, and 3.0 ms after laser emission as produced by simulation in a 5 mm disk diameter of (A) two opposite forces along the Y axis, and (B) a force along Z axis. Applied forces are illustrated by the violet arrows on the most left end side, and direction of the initial central displacement by the arrows in the white circles.%
}}
\end{center}
\end{figure}

To estimate the initial displacement amplitude in this regime, we assumed the medium as homogeneous and isotropic. As the depth of absorption (about 40 $\mu$m) is 100 times smaller than the beam diameter (5 mm), we discarded any boundary effects. The stress, $\sigma_{zz}$, is the sum of the axial strain component and the thermal expansion component \cite{scruby1990laser}:
\begin{equation}
    \sigma_{zz} = (\lambda + 2 \mu) \frac{\partial u_z}{\partial z} - 3(\lambda + \frac{2}{3}\mu) \alpha \frac{ E}{\rho C S \zeta}
    \label{eq:stressThermo}
\end{equation}
where $\lambda$ and $\mu$ are respectively the first and second Lam\selectlanguage{ngerman}é's coefficients, $\alpha$ is the thermal dilatation coefficient, and $\zeta$ is the average depth of absorption. This equation can be simplified by the fact that in most soft media, including biological tissues, $\mu \ll \lambda$. Moreover, in the absence of external constraints normal to the surface, the stress across the surface must be zero, i.e. $\sigma_{zz} (z=0) = 0$. This allows the equation \ref{eq:stressThermo} to be integrated, giving at the surface a displacement $u_z = 3 \alpha E/(\rho C S)$. Substituting the same experimental parameters used previously along with $\alpha$ = 70.10$^{-6}$ K$^{-1}$ (water linear thermal dilatation coefficient), we obtain a displacement $u_z$ of 0.025 $\mu$m. This value is very close to the measured experimental displacement (about 0.02 $\mu$m). Note that both the experimental and theoretical central displacements are directed towards the outside of the medium (see white circle arrow in Figures \ref{figElastoPVA}-(A) and \ref{figGreen}-(A)).

To calculate the propagation of the displacements as shear waves, we must consider the transverse dilatation. Indeed, the illuminated zone is a disk of 5 mm in diameter with a thickness of about 40 $\mu m$ (average depth of absorption along Z axis). The dilatation stress is consequently about two orders of magnitude stronger along the transverse direction than along the Z axis: to compute the displacements along time, we neglected the stress along the Z axis and only considered the stress in the transverse direction. We modeled thus the thermoelastic regime in 2D as two opposite forces directed along the Y axis with a depth of 40 $\mu$m and with an amplitude decreasing linearly respectively from 2.5 to 0 mm and from -2.5 to 0 mm. The magnitude of the force along space and time is stored in a matrix, $H_y^{thermo}(y,z,t)$. Displacements along the Z axis are then equal to the convolution between $H_y^{thermo} (y,z,t)$ and $G_{yz}$ \cite{aki1980quantitative}:
\begin{equation}
G_{yz} = \frac{\cos \theta \sin \theta}{4\pi \rho r} (\frac{1}{c_p^2}  \delta_P - \frac{1}{c_s^2} \delta_S +\frac{3}{r^2} \int\limits_{r/c_p}^{r/c_s}{\tau \delta_\tau d\tau})
\label{eq:Gyz}
\end{equation}
where $\delta_P = \delta(t-\frac{r}{c_p})$, $\delta_S = \delta(t-\frac{r}{c_s})$, $\delta_\tau = \delta(t-\tau)$, $c_p$ and $c_s$ are the compression and shear wave speed respectively, $\tau$ is the time, and $\delta$ is a Dirac distribution. The three terms of the equation correspond respectively to the far-field compression wave, the far-field shear wave, and the near-field component.

Using constants $\rho$ = 1000 kg.m$^{-3}$, $c_p$ = 1500 m.s$^{-1}$ and $c_s$ = 5.5 m.s$^{-1}$ (similarly to experimental measurement), displacement maps along the Z axis were calculated 1.0, 1.5, 2.0, 2.5, and 3.0 ms after force application, as illustrated in Figure \ref{figGreen}-(A). The normalized displacement maps present many similarities to the experimental results displayed in Figure \ref{figElastoPVA}-(A), with an initial central displacement directed towards the outside of the medium, and a propagation of three half cycles.\selectlanguage{english}

Next, we examine the physical characteristics of the other regime observed in our experiments. Solving the equation \ref{eq:eqChaleur} with the same experimental parameters used previously, but with a laser energy of 200 mJ, we find a maximum increase in temperature of 60 K, i.e., a maximum medium temperature of about 360 K assuming a room temperature of about 298 K. While slightly below the vaporization point of our medium, supposed close to 373 K (water vaporization temperature), its proximity to the water vaporization temperature may be sufficient to vaporize the medium. Indeed, it has been demonstrated that graphite and other small particles can act as nucleation sites to facilitate the vaporization of the medium at temperature lower than the vaporization point \cite{Alimpiev_1995}. A series of reactions then leads to displacements inside the medium, which generate shear waves; this constitutes the \textit{ablative regime}.

To estimate the initial displacement amplitude in this regime, we again assume that the medium was homogeneous and isotropic, and we discard any boundary effect. The stress, $\sigma_{zz}$, is now defined as the sum of the axial strain component and a term given by the second law of motion caused by the reaction of the particles ejected outside the medium upon reaching the vaporization point \cite{ready1971effects}:
\begin{equation}  
\sigma_{zz} = (\lambda + 2 \mu) \frac{\partial u_z}{\partial z} - \frac{1}{\rho}\frac{I^2}{(L+C(T_V-T_0))^2}
\label{eq:stressAbla}
\end{equation} where $L$ is the latent heat required to vaporize the solid, and $T_V$ and $T_0$ are the vaporization and initial temperatures, respectively.

By assuming $\mu \ll \lambda$ and a zero stress state at the medium surface, equation \ref{eq:stressAbla} leads to a displacement $u_z = \zeta I^2 / (\rho \lambda(L+C(T_V-T_0))^2)$. Using high-energy experimental parameters, $\zeta \approx \gamma^{-1}$ = 40 $\mu$m (average depth of absorption), $\lambda$ = 2 GPa (first Lam\selectlanguage{ngerman}é's coefficient of water), $L$ = 2.2 MJ.kg$^{-1}$ (vaporization latent heat of water), and $T_V-T_0$ = 373-298 = 75 K, we obtain a displacement $u_z$ of 2.9 $\mu$m. Again, this value is in agreement with the experimentally obtained displacement (2.5 $\mu$m). Both theoretical and experimental displacements are directed towards the inside of the medium (see arrow in the white circle in Figures \ref{figElastoPVA}-(B) and \ref{figGreen}-(B)).

To calculate the propagation of the displacement as a function of space and time, we modeled the ablative regime as a point force directed along the Z axis with a depth of 40 $\mu$m and a constant value from -2.5 to 2.5 mm. The magnitude of the force was stored in a matrix, $H_z^{abla}(y,z,t)$. Displacements along the Z axis are again equal to the convolution between $H_z^{abla}$ and $G_{zz}$ \cite{aki1980quantitative}:
\begin{equation}
G_{zz} = \frac{\cos^2 \theta}{4\pi \rho c_p^2 r}  \delta_P + \frac{\sin^2 \theta}{4\pi \rho c_s^2 r} \delta_S + \frac{3\cos^2 \theta-1}{4\pi \rho r^3} \int\limits_{r/c_p}^{r/c_s}{\tau \delta\tau d\tau}
\label{eq:Gzz}
\end{equation}
with the same notations as presented in equation \ref{eq:Gyz}.

Using the physical quantity values previously defined, displacement maps along the Z axis were calculated 1.0, 1.5, 2.0, 2.5, and 3.0 ms after force application, as illustrated in Figure \ref{figGreen}-(B). The displacement maps present many similarities to the experimental results of Figure \ref{figElastoPVA}-(B), with initial displacement directed towards the inside of the medium, and propagation of only two half cycles.

\begin{figure}[h!]
\begin{center}
\includegraphics[width=1.00\columnwidth]{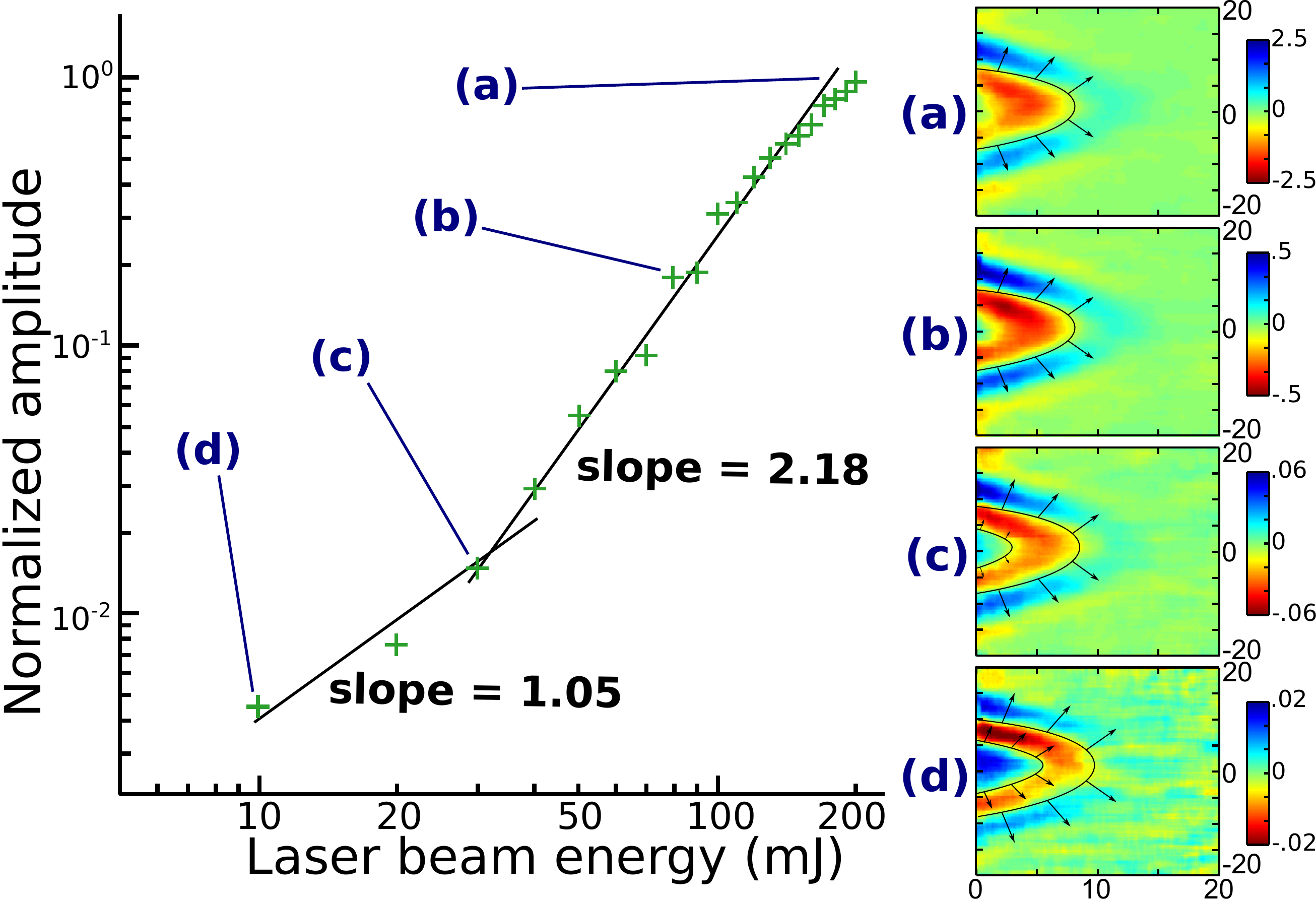}
\caption{{\label{figDepEnergy} Experimentally measured amplitude of the shear wave versus laser beam power (log-log scale). Two linear dependencies are observed: one from 10 to 40 mJ (slope of 1.05), and one from 30 to 200 mJ (slope of 2.18). Four displacement maps at t = 2.0 ms are illustrated, for laser beam energies of 10, 30, 80 and 200 mJ. At low energy, a third half cycle is observed, which disappears when energy increases.%
}}
\end{center}
\end{figure}

Finally, the dependence of the shear wave amplitude versus laser energy was quantitatively investigated by increasing the laser beam energy from 10 to 200 mJ. Amplitudes were averaged over four experiments for each energy level and successive energy levels were randomly chosen to avoid any time-related bias. In addition, impact location was changed after each laser emission to avoid any potential local degradation of the medium. Shear wave amplitude was measured as the mean square amplitude at t = 1 ms of the displacement estimated from ultrasound images between 0 and 10 mm of the medium surface, which was an arbitrary location where shear waves demonstrated high amplitudes. Resulting measurements are illustrated in Figure \ref{figDepEnergy}. Two linear dependencies are observed: one from 10 to 40 mJ, with a slope of 1.05 (R$^2$ = 0.87), and one from 30 to 200 mJ, with a slope of 2.18 (R$^2$ = 0.97). This is in accordance with the theory that displacement is linearly dependent on energy at low energies, i.e. in the thermoelastic regime, but quadratically dependent at high energies, i.e. in the ablative regime. The threshold around 30-40 mJ is specific to our material characteristics, notably its absorption coefficient.\selectlanguage{english}

In the context of shear wave elastography, the thermoelastic regime is a priori preferred over the ablative regime, because it is not destructive. Even if first shear wave elastography experiments assumed that a displacement of a few hundred nanometers would be sufficient \cite{7569924}, displacements of the order of a few micrometers are usually required in practice for ultrasound or magnetic resonance elastography in biological tissues \cite{Nightingale_2001}, \cite{Manduca_2001}. This amplitude is higher than the displacement we observed at 10 mJ (thermoelastic regime), but along the same order of magnitude of the displacement observed at 200 mJ (ablative regime). For application in human body, the lowest fluence (500 J/m$^2$, corresponding to a 10 mJ, 5 mm-diameter laser beam) that was used in these experiments is incidentally 2.5 times above the maximum permissible exposure for skin (200 J/m$^2$) given by the Z136.1-2007 standard of the American National Standard Institute. To overcome this issue, different strategies could be adopted, including using other types of laser, with higher maximum permissible exposure; emission of the laser beam onto a protective absorbing layer, such as a black sheet covering the patient's skin \cite{Li_2014}; or using a higher resolution imaging technique able to track smaller displacements, such as high frequency ($>$100 MHz) ultrasound imaging or optical coherence tomography - in this last case however, low penetration depth may lead to the observation of surface acoustic waves instead of shear waves. Combination with optical coherence tomography may even lead to a real-time, fully remote, small-scale laser-based technique to assess a soft solid stiffness \cite{Li_2011}, \cite{Song_2016a}.

\selectlanguage{english}
\clearpage
\bibliographystyle{plain}

\begin{thebibliography}{10}

\bibitem{aki1980quantitative}
Keiiti Aki and Paul~G Richards.
\newblock {\em {Quantitative Seismology}}.
\newblock Freeman San Francisco, 1980.

\bibitem{Alimpiev_1995}
S.~S. Alimpiev, Ya.~O. Simanovskii, S.~V. Egerev, and A.~E. Pashin.
\newblock {Optoacoustic Detection of Microparticles in Liquids at Laser
  Fluences Below the Optical Breakdown Threshold}.
\newblock {\em Laser Chemistry}, 16(2):63--73, 1995.

\bibitem{Basford_2005}
Alexandra~T. Basford, Jeffrey~R. Basford, Jennifer Kugel, and Richard~L. Ehman.
\newblock {Lorentz-force-induced motion in conductive media}.
\newblock {\em Magnetic Resonance Imaging}, 23(5):647--651, jun 2005.

\bibitem{Catheline_1999}
Stefan Catheline, Francois Wu, and Mathias Fink.
\newblock {A solution to diffraction biases in sonoelasticity: The acoustic
  impulse technique}.
\newblock {\em The Journal of the Acoustical Society of America}, 105(5):2941,
  1999.

\bibitem{17375819}
J~Fromageau, JL~Gennisson, C~Schmitt, RL~Maurice, R~Mongrain, and G~Cloutier.
\newblock {Estimation of polyvinyl alcohol cryogel mechanical properties with
  four ultrasound elastography methods and comparison with gold standard
  testings.}
\newblock {\em IEEE Trans Ultrason Ferroelectr Freq Control}, 54:498--509, Mar
  2007.

\bibitem{Grasland_Mongrain_2014}
P.~Grasland-Mongrain, R.~Souchon, F.~Cartellier, A.~Zorgani, J.~Y. Chapelon,
  C.~Lafon, and S.~Catheline.
\newblock {Imaging of Shear Waves Induced by Lorentz Force in Soft Tissues}.
\newblock {\em Phys. Rev. Lett.}, 113(3), jul 2014.

\bibitem{Grasland_Mongrain_2016}
Pol Grasland-Mongrain, Erika Miller-Jolicoeur, An~Tang, Stefan Catheline, and
  Guy Cloutier.
\newblock {Contactless remote induction of shear waves in soft tissues using a
  transcranial magnetic stimulation device}.
\newblock {\em Physics in Medicine and Biology}, 61(6):2582--2593, mar 2016.

\bibitem{Hirsch_2012}
Sebastian Hirsch, Dieter Klatt, Florian Freimann, Michael Scheel, Jürgen
  Braun, and Ingolf Sack.
\newblock {In vivo measurement of volumetric strain in the human brain induced
  by arterial pulsation and harmonic waves}.
\newblock {\em Magnetic Resonance in Medicine}, 70(3):671--683, sep 2012.

\bibitem{Li_2012}
Chunhui Li, G.~Guan, Z.~Huang, M.~Johnstone, and R.~K. Wang.
\newblock {Noncontact all-optical measurement of corneal elasticity}.
\newblock {\em Optics Letters}, 37(10):1625, may 2012.

\bibitem{Li_2014}
Chunhui Li, Guangying Guan, Fan Zhang, Ghulam Nabi, Ruikang~K. Wang, and
  Zhihong Huang.
\newblock {Laser induced surface acoustic wave combined with phase sensitive
  optical coherence tomography for superficial tissue characterization: a
  solution for practical application}.
\newblock {\em Biomedical Optics Express}, 5(5):1403, apr 2014.

\bibitem{Li_2011}
Chunhui Li, Zhihong Huang, and Ruikang~K. Wang.
\newblock {Elastic properties of soft tissue-mimicking phantoms assessed by
  combined use of laser ultrasonics and low coherence interferometry}.
\newblock {\em Opt. Express}, 19(11):10153, may 2011.

\bibitem{lucas1981iterative}
Bruce~D Lucas and Takeo Kanade.
\newblock {An iterative image registration technique with an application to
  stereo vision.}
\newblock In {\em Proceedings of Imaging Understanding Workshop}, pages
  121--130, 1981.

\bibitem{Manduca_2001}
A.~Manduca, T.E. Oliphant, M.A. Dresner, J.L. Mahowald, S.A. Kruse, E.~Amromin,
  J.P. Felmlee, J.F. Greenleaf, and R.L. Ehman.
\newblock {Magnetic resonance elastography: Non-invasive mapping of tissue
  elasticity}.
\newblock {\em Medical Image Analysis}, 5(4):237--254, dec 2001.

\bibitem{Montalescot_2016}
S.~Montalescot, B.~Roger, A.~Zorgani, R.~Souchon, P.~Grasland-Mongrain,
  R.~Ben~Haj Slama, J.-C. Bera, and S.~Catheline.
\newblock {Electrolysis-induced bubbling in soft solids for elastic-wave
  generation}.
\newblock {\em Appl. Phys. Lett.}, 108(9):094105, feb 2016.

\bibitem{muthupillai1995magnetic}
R~Muthupillai, DJ~Lomas, PJ~Rossman, JF~Greenleaf, A~Manduca, and RL~Ehman.
\newblock {Magnetic resonance elastography by direct visualization of
  propagating acoustic strain waves}.
\newblock {\em Science}, 269(5232):1854--1857, 1995.

\bibitem{7569924}
R~Muthupillai, DJ~Lomas, PJ~Rossman, JF~Greenleaf, A~Manduca, and RL~Ehman.
\newblock {Magnetic resonance elastography by direct visualization of
  propagating acoustic strain waves.}
\newblock {\em Science}, 269:1854--7, Sep 1995.

\bibitem{Nightingale_2001}
Kathryn~R. Nightingale, Mark~L. Palmeri, Roger~W. Nightingale, and Gregg~E.
  Trahey.
\newblock {On the feasibility of remote palpation using acoustic radiation
  force}.
\newblock {\em The Journal of the Acoustical Society of America}, 110(1):625,
  2001.

\bibitem{Plewes_1995}
D.~B. Plewes, I.~Betty, S.~N. Urchuk, and I.~Soutar.
\newblock {Visualizing tissue compliance with {MR} imaging}.
\newblock {\em J. Magn. Reson. Imaging}, 5(6):733--738, nov 1995.

\bibitem{ready1971effects}
John~F Ready.
\newblock {\em {Effects of High Power Laser Radiation}}.
\newblock New York Academic Press, 1971.

\bibitem{scruby1990laser}
Christopher~B Scruby and Leslie~E Drain.
\newblock {\em {Laser Ultrasonics Techniques and Applications}}.
\newblock CRC Press, 1990.

\bibitem{Shan_1993}
Q.~Shan and R.~J. Dewhurst.
\newblock {Surface-breaking fatigue crack detection using laser ultrasound}.
\newblock {\em Appl. Phys. Lett.}, 62(21):2649, 1993.

\bibitem{Song_2016a}
Shaozhen Song, Wei Wei, Bao-Yu Hsieh, Ivan Pelivanov, Tueng~T. Shen, Matthew
  O'Donnell, and Ruikang~K. Wang.
\newblock {Strategies to improve phase-stability of ultrafast swept source
  optical coherence tomography for single shot imaging of transient mechanical
  waves at 16 kHz frame rate}.
\newblock {\em Applied Physics Letters}, 108(19):191104, may 2016.

\bibitem{Xu_2006}
Minghua Xu and Lihong~V. Wang.
\newblock {Photoacoustic imaging in biomedicine}.
\newblock {\em Rev. Sci. Instrum.}, 77(4):041101, 2006.

\bibitem{Zorgani_2015}
Ali Zorgani, R{\'{e}}mi Souchon, Au-Hoang Dinh, Jean-Yves Chapelon, Jean-Michel
  M{\'{e}}nager, Samir Lounis, Olivier Rouvi{\`{e}}re, and Stefan Catheline.
\newblock {Brain palpation from physiological vibrations using {MRI}}.
\newblock {\em Proceedings of the National Academy of Sciences},
  112(42):12917--12921, oct 2015.

\end{thebibliography}

\end{document}